%% file: main.tex
\documentclass[10pt, twocolumn]{article}
\usepackage{arxiv}
\usepackage{geometry}
\providecommand{\keywords}[1]{\textbf{\textit{Keywords:}} #1}
\usepackage{changepage}

\usepackage{authblk}

\input{preamble}

\usepackage{times}

\title{Adaptive probabilistic forecasting of French electricity spot prices}
\author[ \hspace{-1ex}]{Grégoire Dutot$^*$}
\author[1,2,3]{Margaux Zaffran$^*$}
\author[1,4]{Olivier Féron}
\author[1,5]{Yannig Goude}

\affil[1]{Electricité De France R\&D, Palaiseau, France}
\affil[2]{PreMeDICaL project team, INRIA Sophia-Antipolis, Montpellier, France}
\affil[3]{CMAP, École polytechnique, Institut Polytechnique de Paris, Palaiseau, France}
\affil[4]{FiME, Université Paris-Dauphine, Paris, France}
\affil[5]{LMO, Université Paris-Saclay, Orsay, France}

\date{}

\AtBeginDocument{%
  \hypersetup{
    citecolor=orange,
    linkcolor=blindblue}
}

\begin{document}

\maketitle

\def\thefootnote{*}\footnotetext{These authors contributed equally to this work.}
\def\thefootnote{\arabic{footnote}}

\begin{@twocolumnfalse}

\begin{abstract}

Electricity price forecasting (EPF) plays a major role for electricity companies as a fundamental entry for trading decisions or energy management operations. As electricity can not be stored, electricity prices are highly volatile which make EPF a particularly difficult task. This is all the more true when dramatic fortuitous events disrupt the markets. Trading and more generally energy management decisions require risk management tools which are based on probabilistic EPF (PEPF). In this challenging context, we argue in favor of the deployment of highly adaptive black-boxes strategies allowing to turn any forecasts into a robust adaptive predictive interval, such as conformal prediction and online aggregation, as a fundamental last layer of any operational pipeline.

We propose to investigate a novel data set containing the French electricity spot prices during the turbulent 2020-2021 years, and build a new explanatory feature revealing high predictive power, namely the nuclear availability. Benchmarking state-of-the-art PEPF on this data set highlights the difficulty of choosing a given model, as they all behave very differently in practice, and none of them is reliable. However, we propose an adequate conformalisation, \texttt{OSSCP-horizon}, that improves the performances of PEPF methods, even in the most hazardous period of late 2021. Finally, we emphasize that combining it with online aggregation significantly outperforms any other approaches, and should be the preferred pipeline, as it provides trustworthy probabilistic forecasts.

\end{abstract}

\end{@twocolumnfalse}



\section{Introduction}

Electricity price forecasting (EPF) plays a major role for electricity companies as a fundamental entry for trading decisions or energy management operations. As electricity can not be stored, electricity prices  are highly volatile which make EPF a particularly difficult task \citep{weron2014electricity,lago2021forecasting}. 

The increase of renewable production in many countries \citep{rte2022bilan,IEA2022}, the development of storage devices or more generally demand response programs (e.g., electrical vehicle smart charging \citep{Nassar2022}, electric water heater management \citep{AMABILE2021, Marinmoreno2023}) simultaneously entails a need for good EPF and generates more complexity for price modelling. Furthermore, prices can be affected by fortuitous events such as Covid-19 pandemic in 2020-2021 \citep{IEA2021_covid}, the stress corrosion issue which affected French nuclear power plants in 2022 or the crisis of the gas markets triggered by Russia's invasion of Ukraine \citep{IEA2022_outlook}. Trading and more generally energy management decisions require risk management tools which are based on probabilistic EPF \citep{bunn2016analysis}. This supports the advancement of adaptive probabilistic approaches for forecasting prices, which can continuously learn and adjust to the evolving behaviors of EP, resulting in accurate and reliable probabilistic forecasts.

The literature on EPF is growing rapidly and most papers deals with point forecasts \citep{weron2014electricity, lago2021forecasting}. We focus on short term (day-ahead) EPF as the mainstay of short-term power trading in Europe is the day-ahead market. As proposed in \citep{lago2021forecasting}, models used for forecasting electricity prices can be categorized as either statistical, machine learning or hybrid models. 

\textbf{Statistical models} are dominated by  auto-regressive models and their variants, in particular the state ot the art Lasso Estimated AutoRegressive (LEAR) model proposed by \citet{Uniejewski2016} and recently used as state of the art benchmark in \citep{lago2021forecasting, TSCHORA2022}. It consists in a high dimensional ARX model where the fitting process is done by minimizing an elastic net regularization. The high dimension (arround 250 parameters) comes from a large number of lags of prices and forecasts of variable of interests (generation, zonal prices, consumption). As highlighted by \citet{lago2021forecasting} pre-processing of EP such as log transformations or more generally variance stabilizing transformations \citep{Uniejewski2018} are a common practice to deal with heavy tailed distribution. Regarding non-stationarity of the prices, regime switching ARX models are proposed in \citep{Nitka2021}. \citet{Marcjasz1018} propose to average a set of point forecasts obtained from learning with different time windows to derive probabilistic forecasts.

The utilisation of \textbf{machine learning} tools including deep learning approaches  for electricity price forecasting (EPF) has grown over the past decade. Recent studies \citep{TSCHORA2022, jkedrzejewski2022electricity} reveal that complex ML methods such as deep neural networks can achieve better forecasting performances than the LEAR model at the cost of significantly higher computational cost. The relatively important dimension of these models require a significant amount of data for their calibration, making them poor candidate to adapt to abrupt changes in price distribution \citep{Bozlak2024}. \citet{YANG2023} show how graphical neural network could be used to model spatial dependency to forecast the day-ahead electricity prices of the Nord Pool market.

\textbf{Probabilistic price forecasting} is progressively becoming more popular in the forecasting literature following the GEFCom2014 energy forecasting competition \citep{hong2016probabilistic}. This is a natural goal as the final objective EPF is to optimize a financial risk criteria \citep{bjorgan1999financial, DESCHATRE2021}. Most of the previous parametric statistical models are based on statistical assumptions and could be naturally extended to produce probabilistic forecast (more or less accurate as we will explore in this paper). Relaxing distributional assumption, non parametric regression models such as quantile regression have been investigated \citep{UNIEJEWSKI2021105121}. In \citet{Loizidis2024}, machine learning models coupled with boostrap methods are compared with classical time series models for German and Finnish day-ahead market. \citet{marcjasz2023distributional} recently proposed a distributional network that outperforms state-of-the-art benchmarks. \citet{nickelsen2024bayesian} present a Bayesian forecasting framework for the German continuous intraday market and show that orthogonal matching pursuit methods can outperform LEAR. \citet{CORNELL2024} propose quantile regression with varying training-length periods and model averaging to forecast prices of the  South Australia region of the Australian National Electricity Market.

PEPF models face many pitfalls: extreme price spikes, non-stationarity due to exogenous factors inducing time-varying mean and/or volatility. Conformal methods \citep{vovk_machine-learning_1999,papadopoulos_inductive_2002,vovk_algorithmic_2005} and more specifically adaptive conformal methods, proposed for example by \citet{gibbs2021adaptive, zaffran22a}, are a way to adapt PEPF models in a very general way. It can be applied to any of the previously cited PEPFs to improve them. We propose to extend the work of \citet{zaffran22a} to forecast electricity prices in France during the turbulent period 2020-2022. Another framework allowing to adapt PEPF models is online aggregation under expert advice \citep{cesa2006prediction}, which was successfully used in financial non-stationary environments \citep{Remlinger2023, berrisch2024multivariate}. Our aim is to investigate if and how it is possible to make adaptive an existing probabilistic forecasting algorithm. This approach is driven by an operational concern: proposing a plug-in tool that can be applied to any underlying model eases its integration in the current pipeline.

\paragraph{Contributions} 
We list below our main contributions:
\begin{itemize}
    \item \textbf{New data}: we study the recent turbulent period 2020-2022 and we add a new feature, the nuclear availability
    \item \textbf{Benchmark}: we consider state-of-the-art PEPF methods, their windowed versions (rolling window estimation) and benchmark them on this new dataset
    \item \textbf{Analysis} of the improvements (or not) of existing \textbf{online conformal methods}
    \item  Suggestion of \textbf{novel online conformal strategy} coined \texttt{OSSCP-horizon}
    \item Unified framework of \textbf{sequential aggregation} of all these probabilistic forecasting
    \item \textbf{Understanding the benefits} of these 2 frameworks of probabilistic post-processing (i.e. CP and aggregation) and how they can help each other: \emph{sequential aggregation with conformalized expert is the best}
\end{itemize}

\begin{figure*}[!t]
 \centering
 \includegraphics[width=\textwidth]{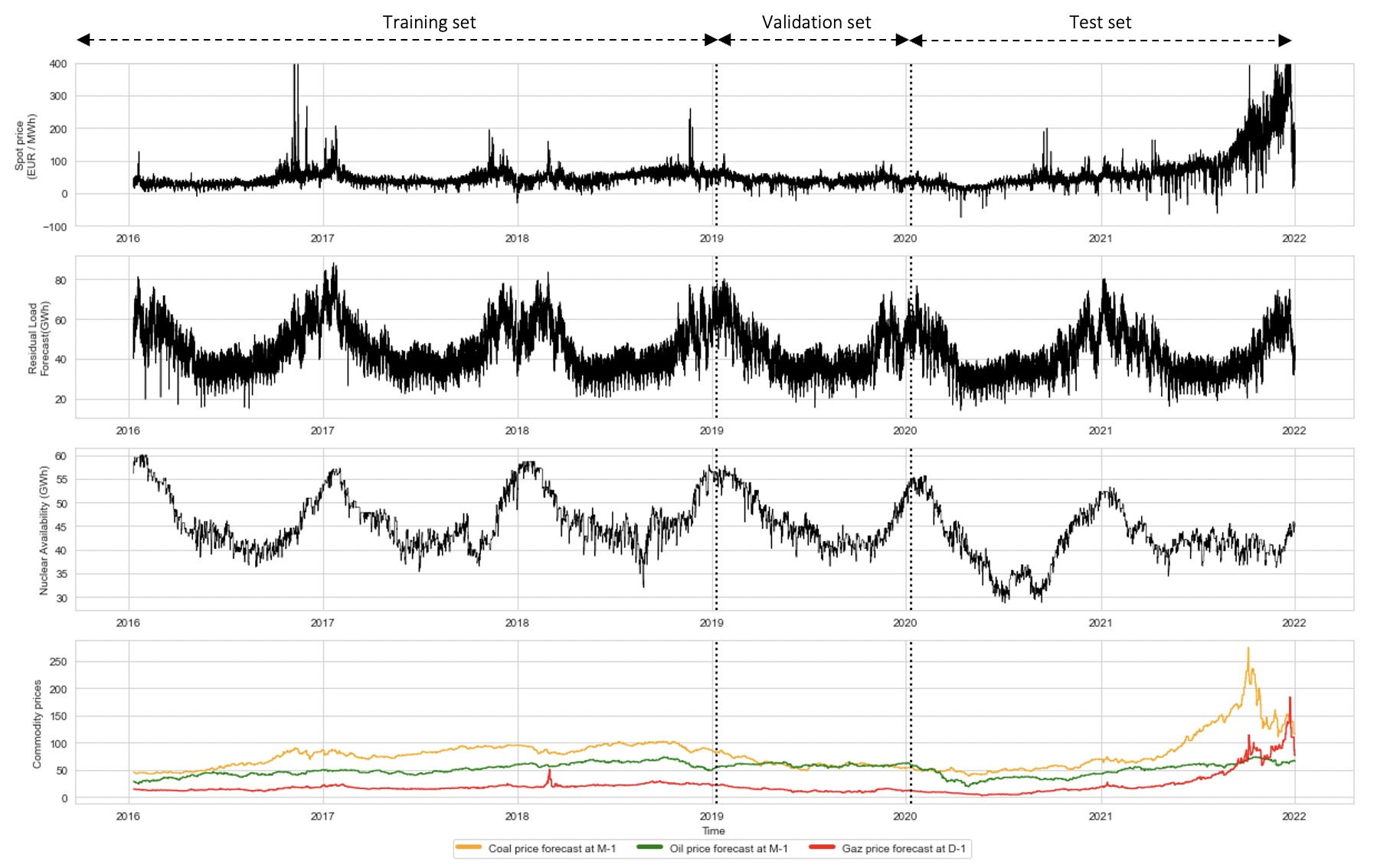}
 \caption{Evolution of the Spot prices (first panel), Residual Load (second panel), Nuclear availability (third panel) and commodity prices (last panel) from 2016 to 2021 ($x$-axis).}
 \label{fig:prices}
\end{figure*}

\section{Data presentation and insightful new explanatory variables}
\label{data}
\subsection{Dataset's description}

The considered dataset spans approximately 6 years of observations at a hourly frequency, from January 11th, 2016 to December 31st, 2021, and is decomposed of a training set (from January 11th, 2016 to December 31st, 2018) to estimate the parameters of the models, a validation test (year 2019) to estimate the hyperparameters, and a test set (years 2020 and 2021) to evaluate the performances (see Figure \ref{fig:prices}). 
We consider the task of forecasting day-ahead (DAH) prices on the French EPEX market. As the 24 hours of day $d$ are fixed from EUPHEMIA\footnote{EUPHEMIA is the algorithm that solves the market coupling problem for the Central West European region, used by EPEX to compute the day-head power prices}'s market clearing at 12:00pm of day $d-1$, the features considered to predict each of them are selected so that they are available before 12:00pm of day $d-1$. More precisely the dataset contains the following features, for a target at day $d$, hour $h$:

\begin{itemize}
    \item the 24 French DAH  prices at days $d-1$ and $d-7$;
    \item the observed daily price of Gas on the French PEG market at $d-1$ and the month-ahead futures prices for Oil (Brent) and Coal (CIF ARA Argus-McCloskey);
    \item the forecasted residual load signal built with data available before 12pm at $d-1$: the load forecasts for the 24 hours of day $d$, estimated on day $d-2$, minus the renewable production forecasts (i.e., wind and solar forecasts estimated on day $d-2$, and the observed run-of-river electricity on \textit{d-2}); 
    \item the availability of French nuclear electricity on day $d$, i.e. the announced available capacity of nuclear generation;
    \item the observed electricity generation from all production types at $d-2$ and $d-7$ (in the case of nuclear energy, the production is divided by the nuclear availability);
    \item the EUR vs. GBP and EUR vs. USD exchange rate (last observed at $d-1$);
    \item the total electricity volume exchanges between France and all its neighbors (observed at $d-2$);
    \item the specific electricity volume exchanges between France and Germany (observed at $d-2$);
    \item dummy variables, including dummy variables for French holidays (as a percentage of the total population concerned), holiday bridges, weekends, and weekdays;
    \item the time of year as a sine and cosine function, as well as a clock variable to capture a possible trend.
\end{itemize}

\subsection{First point forecast and feature importance}

\begin{figure*}[!b]
\centering
\includegraphics[width=0.9\textwidth]{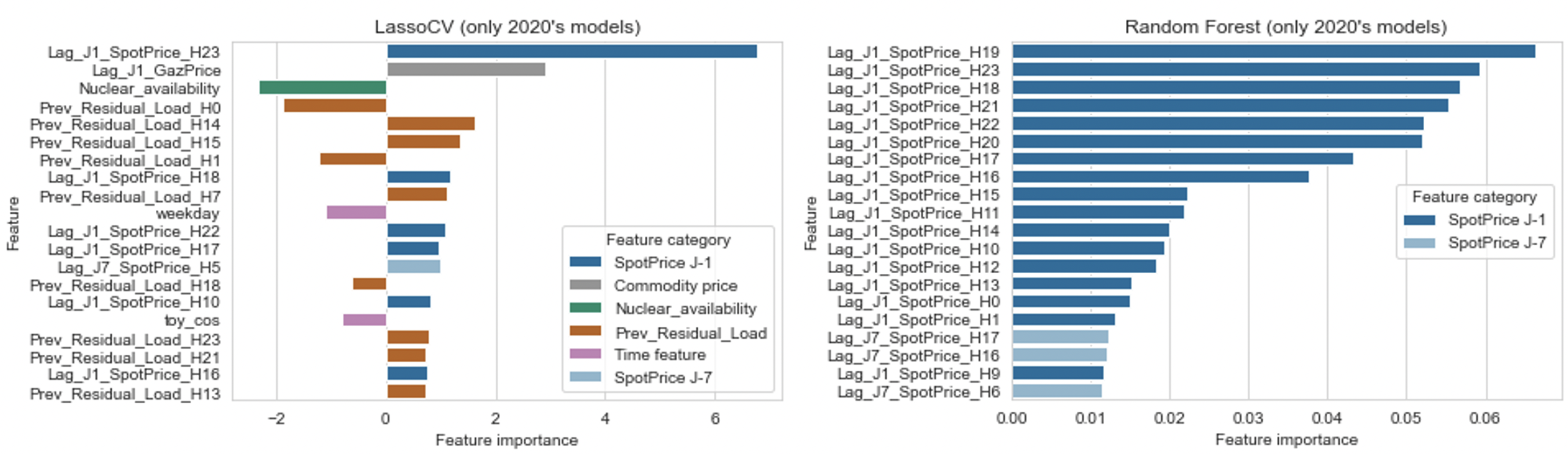}
\caption{Feature ($y$-axis) importance ($x$-axis) for Lasso CV (left panel) and Random Forest (right panel) models. The colors are associated with a type of feature.}
\label{fig:fi_2020}
\end{figure*}

\begin{figure*}[!b]
\centering
\includegraphics[width=0.8\textwidth]{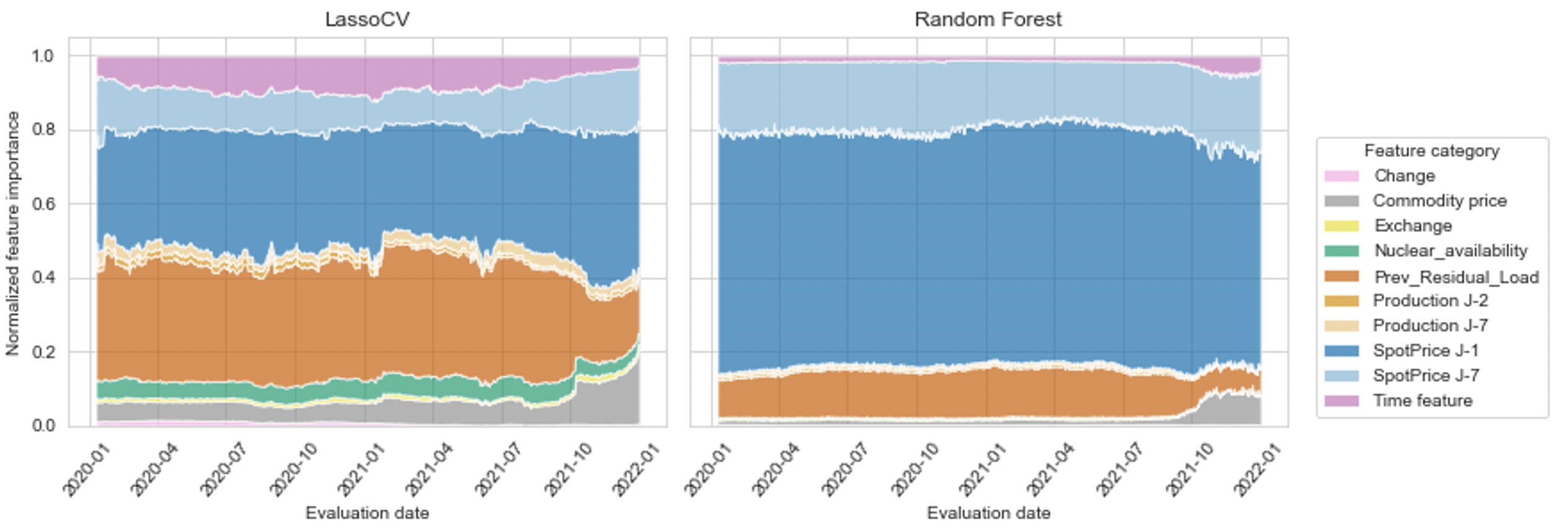}
\caption{Evolution of normalized feature importance ($y$-axis) for Lasso CV (left panel) and Random Forest (right panel) models over the whole test period ($x$-axis). The colors are associated with the features.}
\label{fig:fi_var}
\end{figure*}

The proposed dataset comprises features classically used to forecast electricity prices, and also a new feature, the nuclear availability, for we intuit that nuclear availability has a significant impact on DAH prices due to the French energy mix.

At first we proceed a point forecast exercise, with Lasso CV and Random forest models, to detect the most important features and highlight the relevance of the proposed new variables. Here, the meaning of the term ``feature importance'' varies according to the model: in the case of Lasso CV, it refers to the value of the coefficient associated to a given feature, whereas for Random Forest it refers to the Mean Decrease in Impurity (MDI).

In Figure \ref{fig:fi_2020}, we observe the top 20 mean feature importances over both models trained in 2020. Spot price at H-23 of the previous day is the ``most important'' feature for the Lasso CV model. This is coherent with what is found in \citep{high_exp_power_ld, Ziel_2018}. The MDI-based importances computed for the Random Forest suggests the same conclusion, even though high correlation between all \textit{d-1} spot prices makes the interpretation harder. The Lasso CV model, which allows for a better modelisation with highly correlated features, suggests that gas prices and nuclear availability have a high explanatory power. This speaks in favour of an inclusion of these features in EPF prediction models, at least in the case of the French market.

We also compute the feature importance of both model over every days in the test period and observe the evolution in the predominance of the various feature groups.  To do so, we first aggregate features into groups: ``Change'' for all exchange rates,  ``Commodity price'' for gas, coal and oil prices, ``Exchange'' for all hourly power volumes exchanges, and the rest of features groups are hourly features aggregated at a daily level. The group aggregation consists in summing up the absolute importance value of all features belonging to this group, then normalize these values by the total sum over all groups. Figure \ref{fig:fi_var} represents the evolution we obtain. We observe a considerable change in the relative group's explanatory power: for both the Random Forest and Lasso model, we observe a significant increase in the aggregated explanatory power of the commodity prices, at the expense of the residual load forecast. This indicates an important distribution shift in the relationships between the times series by September 2021. 

\section{Probabilistic forecasting methods}

\paragraph{Notations} Given the nature of the data and in particular the hourly patterns, we will build one model per hour, as explained in \Cref{sec:setting}. From now on, the temporal index $t$ is used and it elapses at a daily rate (i.e., for a given hour $h$). $t = 1$ corresponds to the beginning of the training data, $t = T_0$ marks the end of the training data and $t = T_1$ refers to the last test observation to be predicted. In other words, we aim at predicting the French spot prices between $T_0+1$ and $T_1$, corresponding to the years 2020 and 2021 (see \Cref{fig:prices}).

\subsection{Framework}

One objective of probabilistic forecast is to build \textit{Prediction Intervals} (PIs) for a variable $Y_t$ depending on the covariates $X_t$. Let $\alpha \in [0,1]$ be a \textit{miscoverage rate}. A PI at the $1 - \alpha$ level is expected to contain at least $1 - \alpha$ of the realisations: $\mathds{P}\left(Y_t \in \text{PI}_{1-\alpha}\left(X_t\right)\right) \geq 1-\alpha$, while being as small as possible. In order to retrieve as much information as possible about the distribution of $Y_t$, one can consider multiple values of the miscoverage rate $\alpha$.

A PI can be characterized by two ``point forecasts'': its lower ($\ell(X)$) and upper ($u(X)$) bounds. A natural choice for the PI is $ \ell(X) = Q_{\alpha  / 2}(X)$ and $u(X) =  Q_{1 - \alpha / 2}(X) $, where $Q_\beta$ is the $\beta$-th quantile of the cumulative function distribution (c.d.f.) of the price conditionally to the covariates used to forecast. 

However, in practice, these true $Q$ are never known and we have to estimate them, e.g., using quantile regression \citep{koenker_quantile_2005}. This approach is detailed in \Cref{sec:qr}. 

Another path is to post-process individual predictors (see \Cref{sec:cp}). The individual predictors can either estimate the mean as in point forecasting and the post-processing step will turn them into PI, or directly estimate a conditional quantile (as described in \Cref{sec:qr}).

\subsection{Quantile regression methods}
\label{sec:qr}

We present here the quantile regression methods that we retained for our benchmark study.
These methods were chosen for their good performance on time series data, and in particular on electricity related data. They are all quite easy to fit automatically and have a relatively low computational cost (this is a key asset due to the intensive benchmark including rolling window estimation).

\subsubsection{Description of the methods}

\paragraph{Basics on Quantile  Regression (QR)}

QR \citep{koenker_quantile_2005} replaces the usual quadratic loss by the \textit{pinball loss} to forecast a conditional quantile of the distribution of $Y$ (i.e. the price) given the features~$X$:
\begin{equation*}
\min_{g \in \mathcal{G}} \mathds{E}\left[\rho_\beta (Y - g(X)) | X = x\right],
\end{equation*}
for any $x$, with $\rho_\beta$ the \textit{pinball loss} of level $\beta$: $\rho_\beta  (y - \hat{y}) = (1 - \beta )\vert y - \hat{y} \vert \mathds{1} \{ y \leq \hat{y} \} + \beta \vert y - \hat{y} \vert \mathds{1} \{ y \geq \hat{y} \}$, and $\mathcal{G}$ the class of regressors considered, e.g. linear models, Lasso  (QLR-Lasso), additive non-linear models (QGAM) or gradient boosting regressors (QGB). 

\paragraph{Quantile Linear Regression (Linear QR) and Quantile Lasso (Lasso QR)}
The class of regressors $\mathcal{G}$ is restricted to linear models. For Lasso QR, We perform a Lasso selection process \citep{Lasso90} to deal with the pretty high number of covariates, the class of regressors is thus the linear models on all possible subsets of covariates.

\paragraph{Quantile Generalized Additive Models (QGAM)}
Generalized Additive Models (GAMs) \citep{gam} consists in explaining the conditional expectation $\mu(X)$ of $Y$ over $X$ with a semi-parametric additive structure. The estimation of GAMs is based on a (regularized) mean squared error (MSE) criterion. Our objective is to use GAMs for a QR problem. One could replace the MSE by the pinball loss function in the estimation process as described in the previous paragraph. However, \citet{Fasiolo_2020_qgam} demonstrate that the pinball loss is statistically sub-optimal in this framework and propose a procedure based on the smooth Extended Log-F loss instead. 

\paragraph{Quantile Random Forests (QRF)}

\citet{qrf} adapts Random Forests to the QR task. The same forest is built than for mean-regression, that is a forest grown in order to minimize the mean squared error. However, to adapt to the quantile task at hand, the final decision rule for prediction now corresponds to evaluating an empirical conditional quantile (conditional on the fact that the features of the test point belongs to the corresponding leaves).

\paragraph{Quantile (tree based) Gradient Boosting (QGB)} Gradient boosting machine \citep{friedman2001greedy} are widely used in the forecasting community where it has demonstrated excellent performance for different applications on tabular data \citep{grinsztajn2022tree} or time series \citep{makridakis2022m5}. As for the Random Forests, the regressors are here regression trees. The boosting algorithm consists in adding a sequence of simple models (called weak learners and trained on a subsample randomly selected of the training set) obtained by sequentially fitting a quantile regression tree to the residuals by minimizing the pinball loss, which is a key difference with QRF. 

\subsubsection{Operational pipeline}

We explore these prediction methods through their implementation in the Python package \texttt{scikit-learn} package \citep{scikit-learn} for linear quantile regression, Lasso and QGB. QRF are implemented through \texttt{scikit-garden}. The QGAM are implemented in the \texttt{R} package \citep{qgamR}.

All of these models depend on hyper-parameters, and QGAM additionally requires an exact formula. In particular, we optimized for the regularizer (Lasso), the number of trees and their maximum depth (QRF and QGB), as well as the learning rate and fraction of samples (QGB), and the formula (QGAM). Their estimation is based on grid-searching on the validation set after estimation of mean-regression models on the training set, as illustrated in \Cref{fig:prices}. Therefore, the formula of the QGAM is the same for all quantiles. It includes:
\begin{itemize}
    \item \emph{linear effects}: for the indicator of the week days;
    \item  \emph{univariate non-linear terms}: the announced French nuclear availability, the lagged 2 days of the fossil hard coal and observed nuclear productions, the square root of the lagged one day of the Gaz prices, cosin and sin of the time of year;
    \item \emph{functional smooth effects:} as proposed in \citet{AMARAOUALI20231272} in the context of electricity load forecasting, we model the lagged (one day and one week) prices and the load forecast effects via a functional smooth effect. It allows to capture the effect of these functional (in function of time) covariates over the price at a given instant of the day.
\end{itemize}

\begin{figure*}[!b]
\centering
\includegraphics[width=0.8\textwidth]{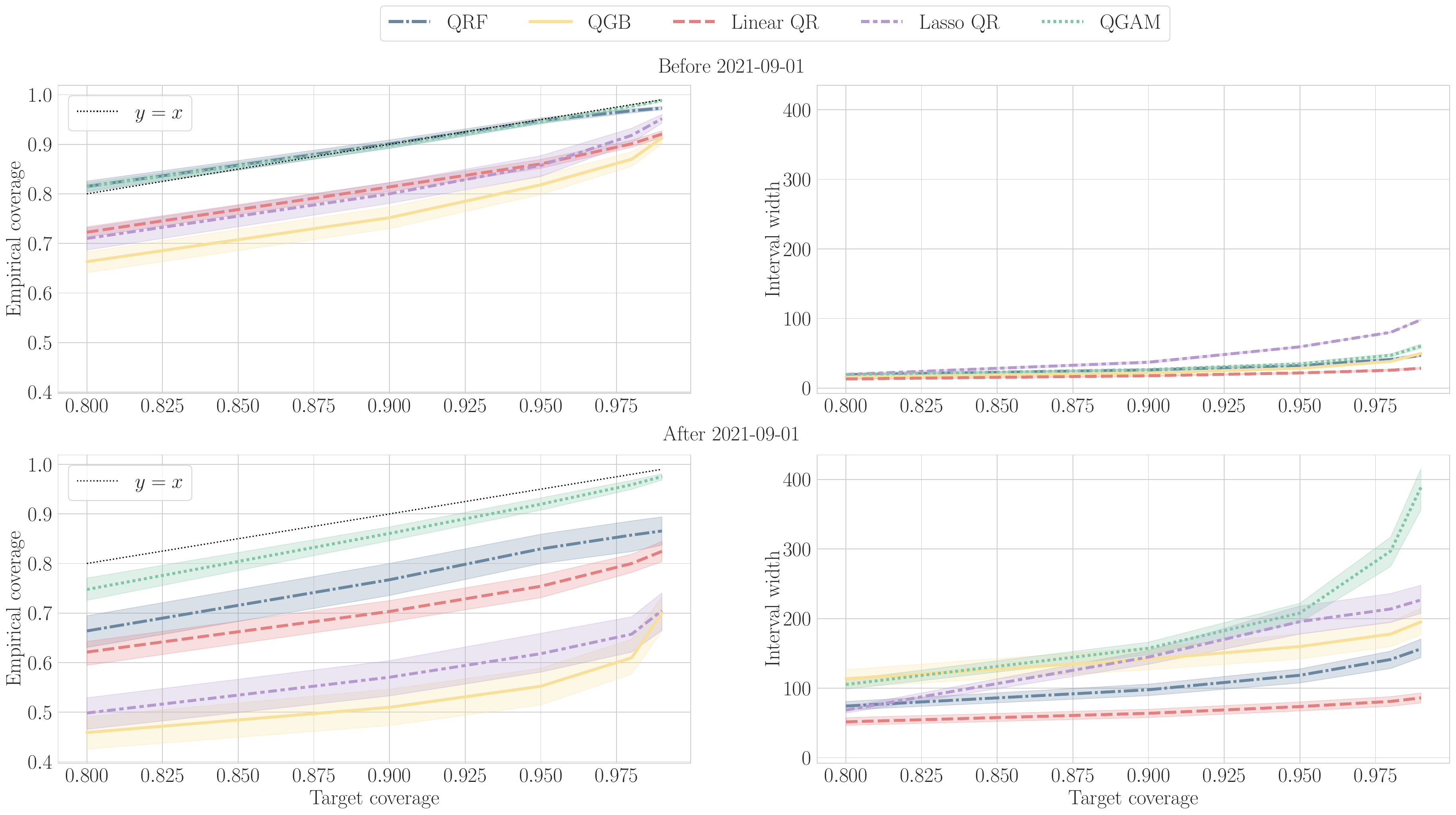}
\caption{PIs's performance of individual probabilistic forecasts at test time, before September 2021 (top row) and after September 2021 (bottom row), for various target coverage levels ($x$-axis). The left column represents the average empirical coverage: the closest to the $y=x$ line the better, and above it is best. The right column represents the average interval width: the lower the better. The colors and shapes are associated with the models. The shaded regions correspond to the 5\% and 95\% empirical quantiles after bootstrapping 500 times the test time series, see \Cref{sec:setting} for details.}
\label{fig:comp_indiv}
\end{figure*}

In this paper we do not consider online re-estimation of the hyperparameters, which in practice is very time consuming and statistically challenging. We study the performance of operational fixed prediction models that can be made adaptive through a plugged-in layer, useful when facing non-stationarity without completely retraining them.

Also, as illustrated in the preliminary results of \Cref{fig:comp_indiv}, before September 2021, only QRF and QGAM achieved \textit{validity}. We explore strategies to recover validity in \Cref{sec:cp}. What is more, none of the probabilistic methods attain the target coverage level after September 2021. Indeed, the high explosion of the prices after this date, both in average and in variability, calls for more adaptive strategies, that we discuss in \Cref{sec:adaptive}. Note that the standard rolling training procedure did adapt to this change as illustrated by the lengths of the PIs after September 2021, but more adaptiveness is required given the strength of the shift and variability.

\subsection{Conformal methods: add-on to traditional probabilistic approaches}
\label{sec:cp}

Conformal Prediction (CP) \citep{vovk_machine-learning_1999,papadopoulos_inductive_2002,vovk_algorithmic_2005} builds PI around any kind of prediction models. These intervals are valid (achieving marginal nominal coverage) in finite samples under the only assumption of exchangeability of the data. Therefore, CP has to be seen as an add-on protective layer to existing probabilistic (or not) forecasts, that is able to robustify them in terms of validity but whose efficiency and shape will always rely on the quality of the underlying forecast.

Suppose that we have $T_0$ random variables $\left\{(X_t, Y_t)\right\}_{t=1}^{T_0}$. For a given miscoverage rate $\alpha \in [0, 1]$, we aim at building a \textit{marginally valid} PI $\widehat{C}_\alpha$ of $Y_{T_0+1}$, i.e. $\widehat{C}_\alpha$ should satisfy:
\begin{equation}
\mathds{P}\left( Y_{T_0+1} \in \widehat{C}_\alpha(X_{T_0+1})\right) \geq 1 - \alpha.
\label{eq:pi_marg}
\end{equation}

To achieve this, Split Conformal Prediction (SCP) \citep{papadopoulos_inductive_2002,lei_distribution-free_2018} randomly splits the $T_0$ data points into a training set $\Tr$ and a calibration set $\Cal$. A regression model $\hat{\mu}$ is then fitted on $\Tr$ and used to predict on $\Cal$ to obtain a set of conformity scores ${\mathcal{S}_{\Cal} = \left\{ S_t := s\left(X_t, Y_t;{\hat\mu}\right),~ t \in \Cal \right\}}$. These scores assess the conformity between the calibration's observed values and the predicted ones: the smaller the better. In the case of regression, they are usually computed using the absolute value of the residuals, i.e. $S_t := s\left(X_t, Y_t;{\hat\mu}\right) = |\hat{\mu}(X_t) - Y_t|$. A corrected\footnote{The correction $1- \tilde{\alpha} = (1-\alpha)(1+\frac{1}{\#\Cal})$ is needed to ensure finite sample validity, because of the inflation of the quantiles.} $(1- \tilde{\alpha})$-th empirical quantile of the conformity scores $Q_{1 - \tilde{\alpha}}(\mathcal{S}_{\Cal})$ is obtained, to finally build the prediction interval $\widehat{C}_\alpha := \left\{y: s(X_{T_0+1}, y; \hat{\mu}) \leq Q_{1 - \tilde{\alpha}}(\mathcal{S}_{\Cal}) \right\}$. In the standard regression case, it boils down to $\widehat{C}_{\alpha}(X_{T_0+1}) = \left[\hat{\mu}(X_{T_0+1}) \pm Q_{1 - \tilde{\alpha}}(\mathcal{S}_{\Cal})\right]$. This procedure is guaranteed theoretically to satisfy \Cref{eq:pi_marg} for any model $\hat\mu$, any sample size $T_0$, as long as the calibration and test data are exchangeable.

Proposed by \citet{cqr}, Conformalized Quantile Regression (CQR) benefits simultaneously from the adaptiveness of classical QR methods and from the theoretical guarantees ensured by CP. Instead of training a mean regression model on the training set $\Tr$, CQR requires to fit two conditional quantile regression models $\hat{q}_{\ell}(\cdot), \hat{q}_{u}(\cdot)$\footnote{Usually $\ell = \alpha/2$ and $u = 1-\alpha/2$, but this is not necessary. \citet{cqr} suggest to choose these values by cross-validation, to improve PI's efficiency.}. In this context, the conformity scores now quantify the error made by the fitted PI $\widehat{C}(x) := [\hat{q}_{\ell}(x), \hat{q}_{u}(x)]$. Precisely, $S_t := s\left(X_t, Y_t;\hat{q}_{\ell},\hat{q}_{u}\right) = \max \left\{\hat{q}_{\ell}(X_t) - Y_t~;~ Y_t - \hat{q}_{u}(X_t) \right\}$.
Accordingly, the PI becomes $\widehat{C}_{\alpha}(X_{T_0+1}) = [\hat{q}_{\ell}(X_{T_0+1}) -Q_{1 - \tilde{\alpha}}(\mathcal{S}_{\Cal}), \hat{q}_{u}(X_{T_0+1}) + Q_{1 - \tilde{\alpha}}(\mathcal{S}_{\Cal})]$.

To account for the temporal aspect of time series, an online and sequential version of SCP is usually considered, in which the split leading to $\Tr$ and $\Cal$ is not random, but constrained so that any point in $\Tr$ occurs before any point in $\Cal$ \citep{pmlr-v128-wisniewski20a,zaffran22a}. See \Cref{fig:splits} for an illustration.

\section{Adaptiveness as a wrapper around individual forecasts}
\label{sec:adaptive}

The online setting---in which the environment reveals the true value before the next prediction---allows to post-process individual predictors to adapt to previous errors (e.g., as done in CP). This approach demonstrates all its interest when stationarity -- and consequently neither exchangeability -- does not hold, as in our case study. One way to implement such a post-processing, coming from the online literature, is online aggregation of predictors, as described in \Cref{sec:agg}\footnote{This does not include Quantile Regression Averaging (QRA) \citep{Nowotarski2014} as it is an offline averaging, thus non-adaptive.}. Another strategy, within the CP framework, is to modify the calibration step of CP (see \Cref{sec:ada_cp}) and make it adaptive.

\subsection{Online aggregation based strategies}
\label{sec:agg}

Adaptive aggregation of \textit{experts}~\citep{cesa2006prediction}, with $K \in \mathds{N}^*$ experts denoted $\left( \hat{f}_t^{(k)} (\cdot)\right)_{k \in \llbracket 1, K \rrbracket}$ being various individual forecasters for the prices at time $t$ (that is a corresponding day $d$ on a given hour $h$) such as the ones introduced in \Cref{sec:qr}, computes an optimal weighted mean of the experts. At each time $t$ (i.e., day $d$, for a given hour $h$), the weights $\omega^{(k)}_t$ assigned to expert $k$ depend on all experts' suffered \textit{losses}, i.e. their performances on the previous time steps until $t-1$. In our case, these performances are evaluated through the pinball loss $\rho_\beta$, standard in quantile regression, with the pinball parameter $\beta$ being the target quantile level. These losses are plugged in the \textit{aggregation rule} $\Phi$, outputting the aggregation weights. Finally, the aggregation rule can include the computation of the gradients of the loss (\textit{gradient trick}, see \citep{cesa2006prediction} for more details). As aggregation rules require bounded experts, a thresholding step is added. Concretely, the aggregated predictor at time $t$, $\hat{f}^{\Phi}_t(\cdot)$, is defined by
\begin{equation*}
\hat{f}^{\Phi}_t(X_t) = \sum_{k = 1}^K \omega^{(k)}_{t} f^{(k)}_{t}(X_t).
\end{equation*}

In our experiments, the different forecasts obtained are aggregated quantile by quantile, using the appropriate pinball loss as a score. The aggregation rule $\Phi$ is set to be the Bernstein Online Aggregation (BOA) \citep{wintenberger2017optimal} algorithm, along with the gradient trick.We use the R package \texttt{OPERA} \citep{opera} to perform such an aggregation, and reorder the quantiles predicted by the aggregation models to avoid quantile crossing.

Recently, \citet{Berrisch_2021} proposed an approach that jointly aggregates every quantile forecasting model together and gives directly a probabilistic prediction as an output, instead of performing independent aggregation for each quantile level. \citet{Berrisch_2021}'s method reduces the number of aggregation parameters to be computed, while yielding preferable probabilistic performances. It is available in the R-Package \texttt{profoc} \citep{profoc_package}, compatible with the BOA method with the gradient trick and automatically reordering the predicted quantiles. It has to be noted that we did not explore the full range of tuning possibilities allowed by this method. In our experiments, both approaches performed similarly. Therefore, to avoid overloading the analysis, we present in this paper only the first method. 

\subsection{Adaptive conformal approaches}
\label{sec:ada_cp}

In addition to online aggregation, we consider another post-processing of individual forecasters which consists in adding a conformal layer on top of them, adaptively. As explained in \Cref{sec:cp}, CP requires exchangeable data, an assumption clearly not satisfied in a time series setting, and even less in our highly non-stationary case study. 

The first theoretically grounded result on CP for dependent data is given by \citet{chernozhukov_exact_2018}: it shows that when the data is strongly mixing and the learned model is close ``enough'' to the underlying data generation process then CP guarantees still hold, along with proposing an extension for full CP\footnote{Full CP is a version of CP that does not require to split the data, at the cost of a bigger computational burden. This is the reason why we do not consider it in this work, along with the fact that full CP can be plugged in on an existing pipeline, making it particularly appealing for operational purposes. The interested reader on full CP can have a look at \citep{vovk_algorithmic_2005}} under which the previous theorem holds. Again, this is not sufficient to encapsulate our setting.  

In practice, Online Sequential Split Conformal Prediction (OSSCP) is often used to take into account the temporal structure, introduced in \citet{pmlr-v128-wisniewski20a, zaffran22a}. The idea is (i) to enforce a sequential split where all the training observations are temporally consecutive, and preceding the ones of the calibration set and (ii) to update this split in order to incorporate the newly observed data points at each prediction step $t+1$, forgiving the oldest ones, leading to adaptive sets $\Tr_t$ and $\Cal_t$. See \Cref{fig:splits} (a) for an illustration. Note that OSSCP does not enjoy any form of theoretical guarantees beyond the exchangeable setting, despite its good empirical performances in the time series framework, as highlighted in \citep{zaffran22a}.

\subsubsection{Improving CP online adaptiveness: \texttt{OSSCP-horizon}}

One drawback of OSSCP is that the set on which the models were fitted can be far from the points on which it will be applied (either calibration or test points). If the temporal data suffers from a strong distribution shift, this may hinder the accuracy of the base learner, and therefore the performances of the PI, both in terms of coverage (the exchangeability assumption is not satisfied anymore) and in terms of efficiency, i.e. interval's length (as large errors cause large intervals).

\begin{figure}
\centering
\includegraphics[width=\linewidth]{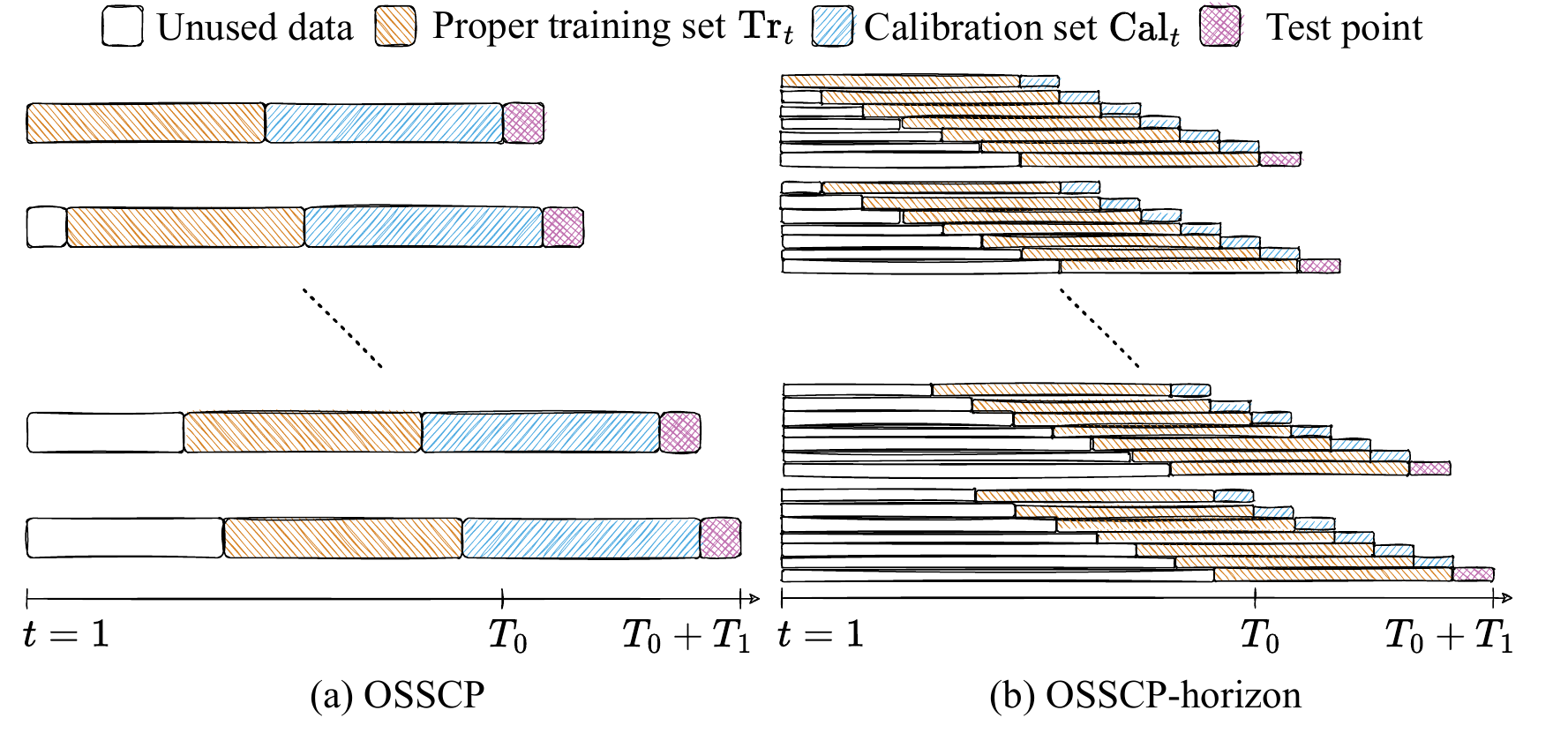}
\caption{Scheme of OSSCP (a) and our proposal (b), \texttt{OSSCP-horizon}, when the horizon is 1.}
\label{fig:splits}
\end{figure}

In order to avoid high errors on the calibration and test points, we propose a new approach, coined \texttt{OSSCP-horizon}. The idea is to ensure that the underlying model is trained on the data just preceding each calibration point: in other words, to only compute test errors of horizon one, as is the forecast horizon. More generally, for any forecasting task at horizon $h$, \texttt{OSSCP-horizon} computes calibration errors of horizon $h$. See \Cref{fig:splits} (b) for an illustration. Formally, at prediction time $T+1$, \texttt{OSSCP-horizon} thus builds the calibration set as follows:

\begin{itemize}
    \item For each $X_t \in \Cal_T$, fit quantile regression estimators $\hat{q}_{\ell}^{-(t)}$, $\hat{q}_{u}^{-(t)}$ on $\left\{\left( X_{t - |\Tr|}, Y_{t - |\Tr|} \right), \ldots, \left( X_{t - 1}, Y_{t - 1} \right) \right\}$\footnote{For a horizon $h \neq 1$, then $\hat{q}_{\ell}^{-(t)}$, $\hat{q}_{u}^{-(t)}$ are fitted on $\left\{\left( X_{t - |\Tr|}, Y_{t - |\Tr|} \right), \ldots, \left( X_{t - h}, Y_{t - h} \right) \right\}$.};
    \item Compute the calibration score $S_t = s\left(X_t, Y_t;\hat{q}_{\ell}^{-(t)},\hat{q}_{u}^{-(t)}\right)$ and add it to the set of scores $\mathcal{S}_{\Cal_T}$.
\end{itemize}
After having built $\mathcal{S}_{\Cal_T} = \{ S_{T - |\Cal| + 1}, \ldots, s_{T}\}$, \texttt{OSSCP-horizon} computes the PI for the test point $X_{T+1}$: 
\begin{align*}
\widehat{C}_{\alpha}(X_{T+1}) := & \left[ \hat{q}_{\ell}^{-(T+1)}\left(X_{T+1}\right) - Q_{1 - \tilde{\alpha}}\left(\mathcal{S}_{\Cal_T}\right); \right. \\ 
& \left. \;\; \hat{q}_{u}^{-(T+1)}\left(X_{T+1}\right) + Q_{1 - \tilde{\alpha}}\left(\mathcal{S}_{\Cal_T}\right) \right].
\end{align*}

Again, while demonstrating empirical improvements upon standard OSSCP in the temporal setting, \texttt{OSSCP-horizon} does not enjoy any form of theoretical guarantees. To theoretically account for the online setting, a popular method is Adaptive Conformal Inference (ACI) \citep{gibbs2021adaptive}.

\subsubsection{Adaptive Conformal Inference (ACI)} 

Proposed in \citep{gibbs_adaptive_2021}, ACI adapts CP to an arbitrary online setting, including temporal distribution shits. To do so, ACI recursively updates the \textit{effective} miscoverage rate $\tilde{\alpha} := \alpha_t$ used in the computation of the PI. Set $\alpha_{1} = \alpha$. For $t \geq T_0$, and for a chosen $\gamma \geq 0$ the ACI update formula is:

$$\left\{
    \begin{array}{ll}
        \widehat{C}_{\alpha_t}(X_{t})  := [\hat{q}_{\ell}(X_{t}) - Q_{1 - \alpha_t}(\mathcal{S}_{\Cal_t}), \hat{q}_{u}(X_{t}) + Q_{1 - \alpha_t}(\mathcal{S}_{\Cal_t})]\\
        \alpha_{t+1} = \alpha_{t} + \gamma \left( \alpha - \mathds{1}_{ \{Y_t \not\in \widehat{C}_{\alpha_t}(X_{t})  \}}\right)
    \end{array}.
\right.$$ 

The underlying idea is the following. If the PI does not cover at time $t$, then $\alpha_{t+1} \leq \alpha_{t}$ which increases the size of the PI. Conversely, the size of the interval decreases gently at time $t+1$ when it covers at time $t$. As noted in \citep{zaffran22a}, it is possible to have $\alpha_t \geq 1$ or $\alpha_t \leq 0$: the former case is quite rare and produces by convention $\widehat{C}_{\alpha_t} = \left[\hat{q}_{\ell}(\cdot), \hat{q}_{u}(\cdot)\right]$; however, the latter can happen frequently, especially for a high $\gamma$, giving a prediction interval of infinite size ($\widehat{C}_{\alpha_t} \equiv \mathds{R}$).

The main theoretical result on ACI is that for any sequence $(X_t, Y_t)_t$, $\left\vert \frac{1}{T_1-T_0} \sum\limits_{t=T_0+1}^{T_1} \mathds{1}\left\{ y_t \in \widehat{C}_{\alpha_t}(X_t)  \right\} - (1-\alpha) \right\vert \leq \frac{2}{\gamma (T_1-T_0)}$. It shows the asymptotically valid frequency of ACI intervals for any arbitrary (possibly adversarial) distribution. 

Note that the convergence rate is in $\gamma^{-1}$, hence favoring large $\gamma$ which are the ones leading to more variability and in the extreme case to infinite PIs (discussed previously). This illustrates the need for guidance on how to choose properly $\gamma$, and even avoid having to choose it and being able to switch between different $\gamma$ depending on the current data distribution's evolution.

\subsubsection{AgACI}

The goal of AgACI, proposed in \citep{zaffran22a}, is precisely to provide a parameter-free method based on ACI, that can adapt to temporal changes in the data distribution adaptively. Given a list of $K$ $\gamma$ values $\left\{\gamma_k\right\}_{k=1}^K$, AgACI works as an adaptive aggregation of experts \citep{cesa2006prediction} (see also \Cref{sec:agg}), with expert $k$ being ACI with parameter $\gamma_k$. At each prediction step $t$, it performs two independent aggregations of the $K$ ACI intervals $\widehat{C}_{\alpha_{t,k}}(\cdot) \overset{\text{not.}}{=} [\hat{b}^{(\ell)}_{t,k}(\cdot), \hat{b}^{(u)}_{t,k}(\cdot)]$,
one for each bound, and outputs $\widetilde{C}_{t}(\cdot) \overset{\text{not.}}{=} [\tilde{b}^{(\ell)}_{t}(\cdot), \tilde{b}^{(u)}_{t}(\cdot)]$. According to \citet{zaffran22a}, the standard different aggregation rules gave similar results. In this work, we restrict ourselves to the setting of \citep{zaffran22a}, that is BOA, with the gradient trick. 

\subsubsection{Latest related works}
\label{sec:rel_works}

Since the analysis presented in this paper was performed, the line of research on adaptive and online conformal approaches has been expanding fast. Recent developments include: \citet{gibbs2023conformal} improving on ACI by online aggregation on a grid of different $\gamma$, similarly to AgACI, at the crucial difference that the aggregation is on the value of $\alpha_t$ and not on the lower and upper bounds independently (\Cref{sec:results} highlights why we argue in favor of different aggregations); \citet{bastani2022practical} who achieve stronger coverage guarantees (conditional on the effective level, and conditional on specified subsets of the explanatory variables); \citet{pmlr-v202-bhatnagar23a} enjoy anytime regret bound, by leveraging tools from the strongly adaptive regret minimization literature; \citet{angelopoulos2023conformal} who extend upon ACI ideas by relying on control theory to add more information on the temporal structure; \citet{angelopoulos2024online} proposing to use adaptive learning rates $\gamma_t$ in ACI. 

Our goal in this analysis is to deeply investigate the improvements, or not, brought by conformal as one of the layers for probabilistic forecasts with an operational lens. Therefore, we restricted the study to OSSCP, \texttt{OSSCP-horizon}, and AgACI as it has already shown benefits on electricity prices and does not require to select any hyper-parameter \citep{zaffran22a}. Indeed, it allows us to easily understand what is the cause of the improved or declined performance. Furthermore, the most recent works are either complex structures (thus less interpretable) or depend on hyper-parameter tuning, making them more costly to implement in operational use. 

\section{Application and results}

\subsection{Setting and evaluation}
\label{sec:setting}

\paragraph{Experimental details} In order to span a wide range of the price distribution function, we vary the PIs' miscoverage level $1 - \alpha > 0.6$. For the final probabilistic forecasts, the overall training set comprises 4 years of data, from 2016 to 2019 included (i.e. merging the training and validation sets).

Due to training time constraints, we trained and evaluated the considered models on hours 3, 8, 13, 18, and 23 of every day. These 5 hours encompass best the different phases of hourly electricity prices in a given day, while uniformly covering the 24 hours of the day.

Finally, due to the high non-stationarity, we trained each of the base models presented in \Cref{sec:qr} on different window sizes: approximately 4 years, 3 years, 2 years, 1 year, 270 days, 180 days, and 90 days. For the sake of clarity, for each analysis performed, the largest window size will be selected and presented in this paper. In the same vein, the calibration size of the conformal approaches (\Cref{sec:cp,sec:ada_cp}) varies among 25\%, 50\% and 75\% of the overall windowed training set. Again, to ease interpretation of our results, we present here only the results for a calibration set of proportion 50\% (except if stated otherwise) as it allows for an intermediary adaptation speed, hence being a good trade-off between up-to-date quantile regression models and calibration set large enough to perform the estimation of the highly non-stationary conformal correction. We recall that in the i.i.d. setting a general rule of thumb for the calibration size is around 25\% \citep{Sesia2020}. In our study, the impact of non-stationarity induces a need for a trade-off between adaptivity and the calibration window length.   

\begin{figure*}[!b]
 \centering
 \includegraphics[width=0.8\textwidth]{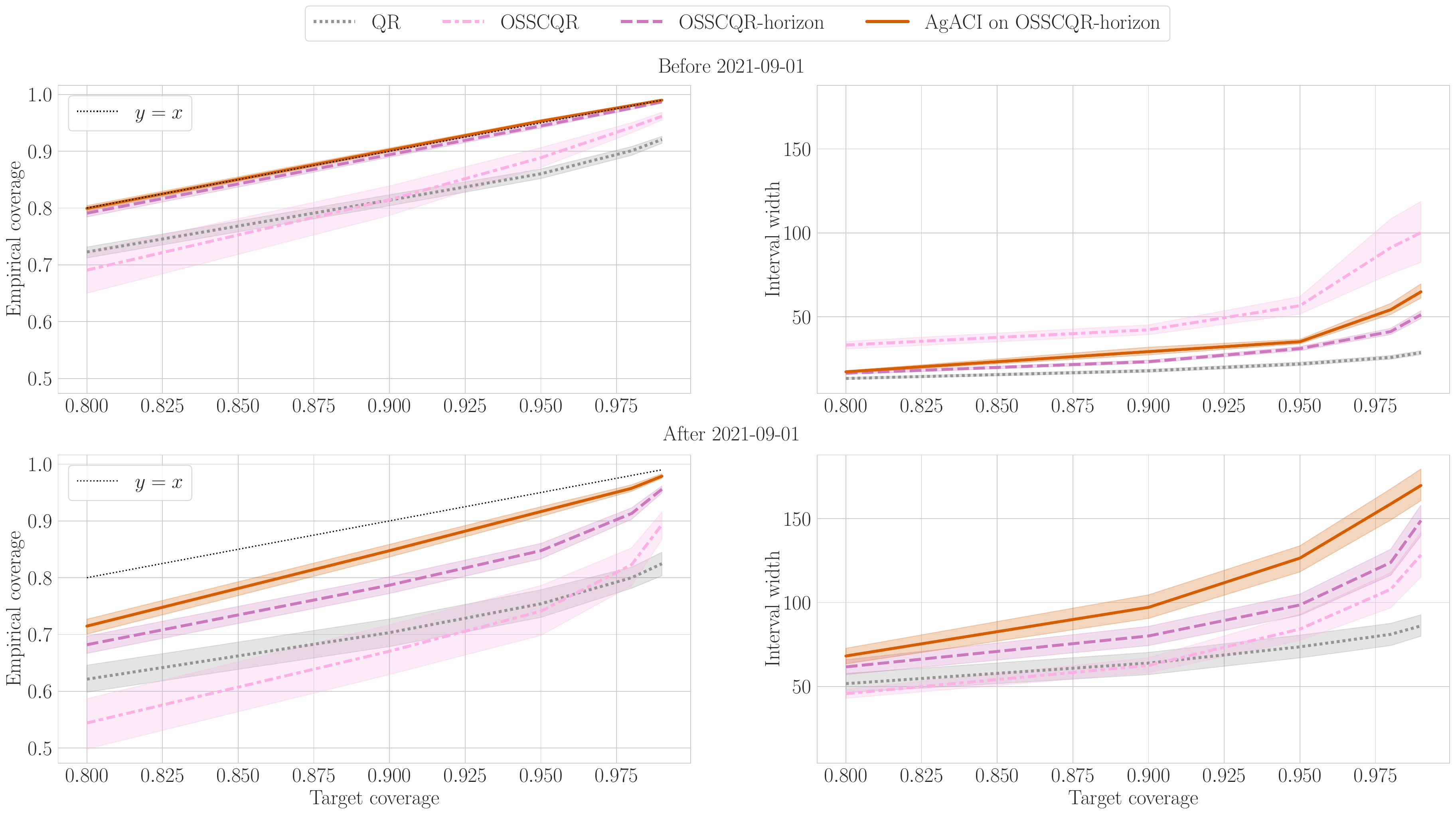}
 \caption{PIs's performances with different levels of conformalisation on the \textbf{quantile linear model}, before September 2021 (top row) and after September 2021 (bottom row), for various target coverage levels ($x$-axis). The colors and shapes are associated with the conformalisation layers. The shaded regions correspond to the 5\% and 95\% empirical quantiles after bootstrapping 500 times the test time series.}
 \label{fig:linear}
\end{figure*}

\begin{figure*}[!b]
\centering
\includegraphics[width=0.8\textwidth]{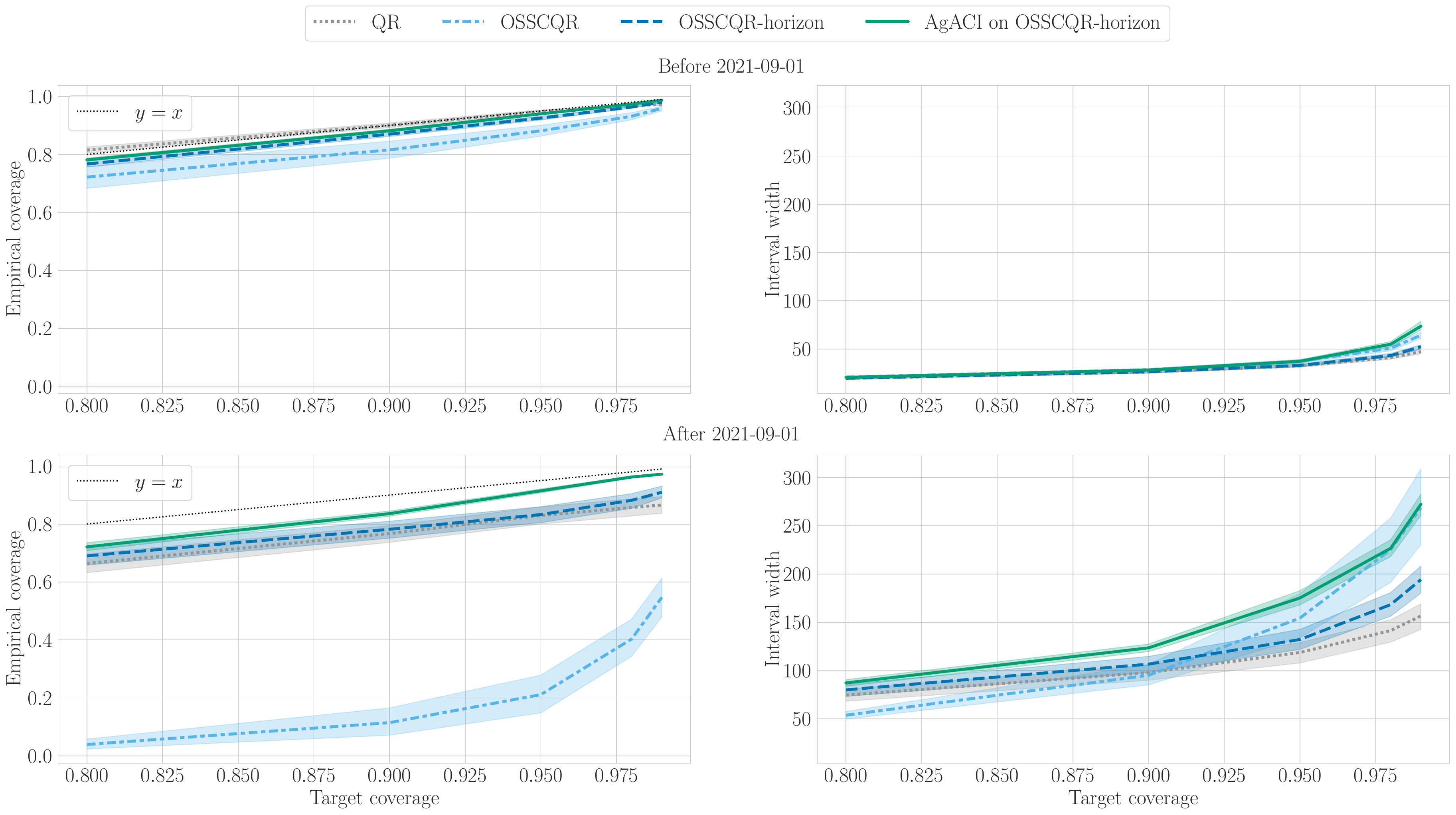}
\caption{Same caption than \Cref{fig:linear} but for the \textbf{quantile random forest model}.}
\label{fig:qrf}
\end{figure*}

\paragraph{Evaluation procedure} The main challenge of evaluating a probabilistic forecast is that the true distribution of the underlying process cannot be observed. Hence, it is impossible to compare the estimated distribution with the actual distribution of the true spot prices. This is not the case for a sequence of PIs $\left( \left[\hat{b}^{(\ell)}(\cdot),\hat{b}^{(u)}(\cdot)\right] \right)_t$ that can be evaluated through:
\begin{itemize}
    \item \textbf{empirical average coverage}, \\$\frac{1}{T_1-T_0} \sum\limits_{t=T_0+1}^{T_1} \mathds{1}\left\{ y_t \in \left[\hat{b}^{(\ell)}(x_t),\hat{b}^{(u)}(x_t)\right] \right\}$, that should be close and above to the target level $1-\alpha$ for \textit{validity} (also known as \textit{reliability}),
    \item \textbf{empirical average length}, $\frac{1}{T_1-T_0} \sum\limits_{t=T_0+1}^{T_1} \hat{b}^{(u)}(x_t) - \hat{b}^{(\ell)}(x_t)$, for \textit{efficiency}\footnote{Indeed, achieving exactly $1-\alpha$ coverage can be trivially done by outputting $1-\alpha$ of the time $\mathds{R}$ and the empty set otherwise, which is critically uninformative. Thus, one wants to attain \textit{validity} while minimizing the size of the resulting intervals, that is maximizing \textit{efficiency}.} (also known as \textit{sharpness}).
\end{itemize}

For each of these metrics, confidence intervals are constructed by time series bootstrapping (non-overlapping moving block bootstrap) \citep{kunsch1989jackknife, politis1994stationary}.

Results on the CRPS are provided in \ref{app:crps}. Indeed, our goal is really to compare PIs and not predictive distributions. Therefore, the forecasts' objective is truly to \textbf{be as sharp as possible while satisfying validity}.

\subsection{Results}
\label{sec:results}

\paragraph{Impact of the conformalisations} In \Cref{fig:linear,fig:qrf} we represent the performance of Linear Quantile Regression and Quantile Random Forest respectively, with various layers of conformalisation. The display choice of these two base models is motivated by the fact that they represent a diverse range of modelisation. 

In both cases, we observe that a naive conformalisation -- in the form of OSSCP -- does not allow to achieve the nominal coverage level, neither before nor after September 2021. 

Yet, our proposal \texttt{OSSCP-horizon} does improve drastically the coverage level: before September 2021 it manages to reach the target level while improving the lengths of the PIs, and after September 2021 it allows to reduce the gap with the target considerably (linear model), while recovering the approximatively satisfactory performances of the individual QRF that was deteriorated by OSSCP. 

Finally, making the conformalisation even more adaptive through the use of AgACI especially enhances validity after September 2021. Yet, it has to be noted that it seems to be insufficiently adaptive to perfectly reach the target level.

\begin{figure*}[!b]
    \centering
    \includegraphics[width=0.8\textwidth]{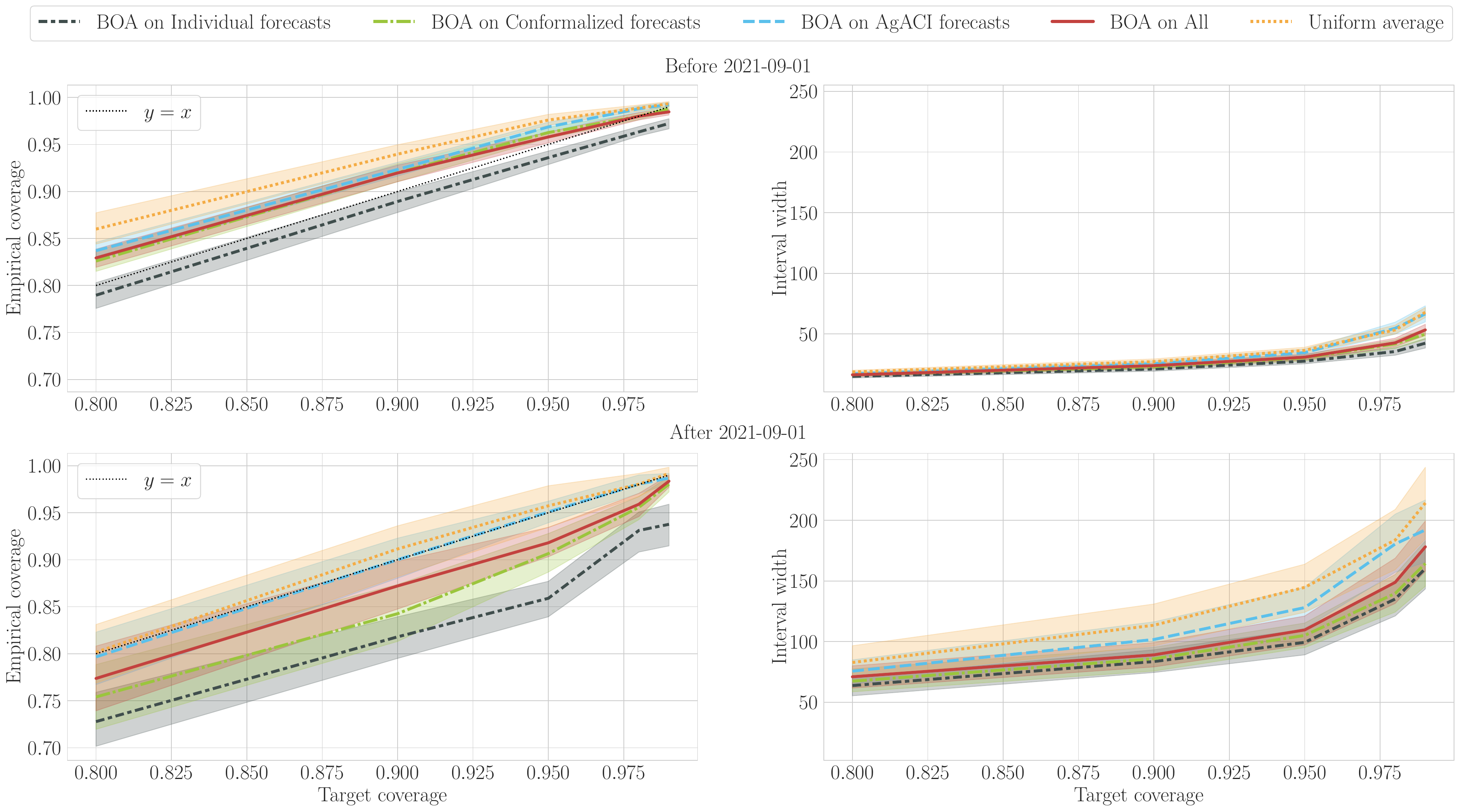}
    \caption{PIs's performances of online aggregation on multiple set of experts with windowing, before September 2021 (top row) and after September 2021 (bottom row), for various target coverage levels ($x$-axis). The colors and shapes are associated with the set of experts. The shaded regions correspond to the 5\% and 95\% empirical quantiles after bootstrapping 500 times the test time series.}
    \label{fig:aggs}
   \end{figure*}

\paragraph{Analysis of various aggregations} Therefore, we go further and add another adaptive post-processing layer by performing online aggregagation. In \Cref{fig:aggs} we compare the performances of various aggregations, each of them considering a different set of experts (individual forecasts, \texttt{OSSSCP-horizon} forecasts, AgACI forecasts, and all of them). As a baseline, we add the uniform average of all of these experts. For each of the aggregation, we compared aggregating forecasts with a unique window size for training with aggregating forecasts with multiple training window size (hence augmenting the number of experts in the set). This latter strategy is usually referred to as \textit{windowing} \citep{Marcjasz1018}. We selected the best aggregation (namely aggregating AgACI forecasts with windowing) and, for the sake of readability and for coherence, we displayed in \Cref{fig:aggs} all the aggregations with windowing. It has to be noted that there is a lot of variability, as it can be seen in \Cref{fig:aggs}, and that for some aggregation the best choice was in fact without windowing.

\Cref{fig:aggs} highlights that online aggregation improves considerably the robustness to non-stationarity in terms of validity. Furthermore, after September 2021, online aggregation on AgACI forecasts enhances the sharpness of the forecasts with respect to the uniform average, that has similar coverage. This can be explained by the fact that the individual performances degrade in this non-stationary environment, leading to aggregation's weights close to uniform so as to minimise the risk (as we will also see in the next analysis).

\begin{figure*}[!t]
    \centering
    \includegraphics[width=\textwidth]{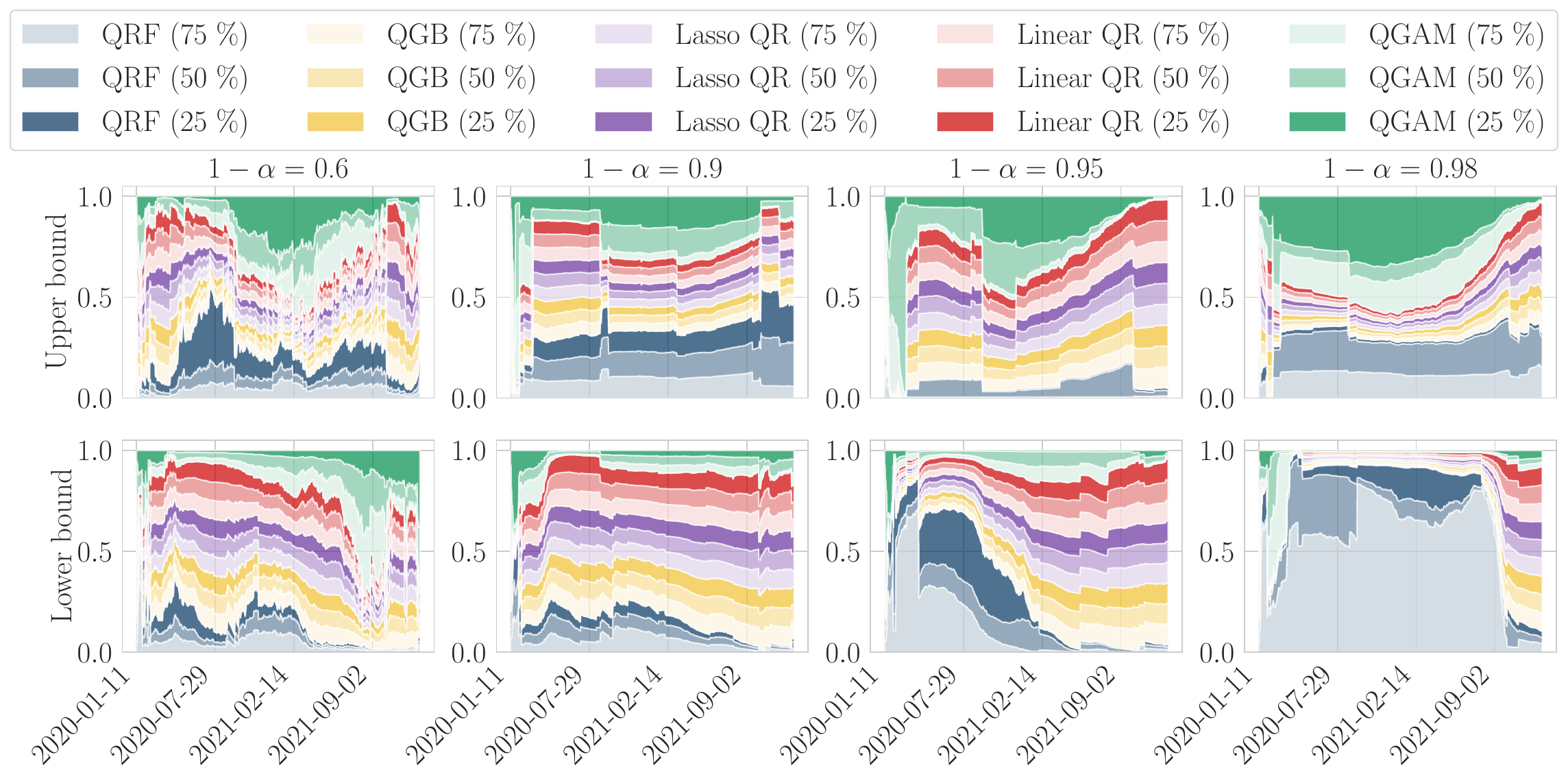}
    \caption{Temporal evolution ($x$-axis) of the weights associated with each expert in the online aggregation, for different values of (columns). The top row (resp. bottom row) shows the weights assigned for the upper (resp. lower) bound forecast. The colors correspond to the base model on which AgACI is applied to, and the transparency to the proportion of training data kept for actually fitting these base models.}
    \label{fig:weights}
   \end{figure*}

\paragraph{Analysis of aggregation of various AgACI: applying the best conformalisation possible (AgACI) on each model and then aggregating them} In \Cref{fig:weights} we represent the evolution of the weights associated to each of the AgACI (the color representing the base model, and the shade of it indicating the calibration percentage) with time $x$-axis, for various coverage level (columns). To improve readability, we display these weights for the aggregation without windowing. 

The first striking observation is the presence of temporal ruptures in the weights' distribution. They are informative as they are associated with domain phenomena, which depend on the considered bound (lower or upper). Particularly, the first one happening is the big negative spike in Easter 2020 (April 13, 2020, see top row of \Cref{fig:prices}) due to both the public holiday and the Covid-19 lockdown. This especially affects the lower bound. The second one occurs in the second fortnight of September 2020 when the first extreme positive peaks take place, impacting the upper bound. These positive spikes are mainly due to a very low wind generation in France (less than 1 GW) and more generally in Europe, along with a French nuclear production well below its level of previous years at the same time. The last significant rupture is around October 2021, when spot prices start to rise drastically and get more and more volatile, corresponding to the increase in level and volatility of gas and carbon emission prices. This one affects both the lower and upper bounds. In particular, the weights' distribution becomes uniform after this rupture, which is expected in a setting where the aggregation tries to minimize the risk with experts performing poorly.

The second observation is that the methods on which the aggregation places the most of the weights is different depending on the bound: remarkably, at the levels 0.95 and 0.98, the lower bound places high mass on quantile random forests, while the upper bound relies more on qgam. This can be explained by the fact that the various methods depend differently on the provided features: additive models such as qgam or linear ones have a great extrapolation ability, while random forests and gradient boosting benefit from more flexibility on features' interaction modeling. This idea is also reflected in \Cref{fig:fi_2020,fig:fi_var} comparing the feature importance in Lasso with the one of Random forest. 

Lastly, for high levels of coverage such as 0.95 and 0.98, the aggregation also places weights on different training size depending on the bound. While the upper bound favors small training size, the lower bound encourages large training size. This might be due to the effective sample size which is required to appropriately learn the lower quantiles of the prices, which are less impacted by the non-stationarity; while the upper bound is particularly complex to model, and having more data points correct the predictive model through conformalisation might be a better usage of the available data.

These three key observations argue in favor aggregating independently the upper and lower bounds. 

\section{Conclusion and perspectives}

In this study, we have analysed the performances of a wide range of probabilistic methods in a particularly challenging task: forecasting electricity spot prices in France in 2020 and 2021. On the design, we have highlighted the importance of including the new explanatory variable corresponding to the nuclear plants' availability. We were also able to bring new insights into the post-processing of individual forecasts, such as conformalisation or aggregation. Indeed, our extensive experiments demonstrate that $i)$ conformalisation, when appropriately done as through \texttt{OSSCP-horizon}, considerably improves PI's quality despite the non-stationarity, $ii)$ online aggregation of experts is extremely powerful in terms of adaptiveness bringing enhanced PI's performances and taking advantage of windowing, $iii)$ combining both conformalisation and online aggregation appears on this data set to be the best strategy, and most importantly sheds light on many domain phenomena thanks to great interpretability.

There are many avenues for future works. From the electricity lens, the prices have continued to evolve significantly since 2022 and pursuing the study on newer data would undoubtedly yield new knowledge. Speaking of which, our study did not investigate the crucial question of peaks and extreme forecasts, dominant in electricity prices. Works on online procedure tailored for extremes have already been deployed \citep{himych:hal-04460731}, and it might be relevant to see how it can be paired with conformal approaches. Another natural perspective that would deepen our understanding on the benefits of conformalisation is to conformalize the aggregated models as suggested in \citet{susmann:hal-04539380}, as opposed to aggregating the conformalized models which is what we performed. It would also be interesting to assess the performances of the most recent online conformal algorithms (listed in \Cref{sec:rel_works}), that might be better suited for non-stationarity. Finally, our angle of approach is to showcase the advantages of black-box plugs-in such as CP and aggregation. It is attractive to couple it with recent developments that enhance the interpretability of complex statistical models, such as \citet{Wood2022}.

\bibliographystyle{plainnat}
\bibliography{bibli}

\newpage

\appendix

\section{Results on the CRPS}
\label{app:crps}

To assess the performance of a probabilistic method on the overall range of quantiles, one can use the Continuous Ranked Probability Score (CRPS). This score is originally described in terms of the predictive CDS $\hat{F}_{d,h}$ :
$$CRPS(\hat{F}_{d,h}, y_{d, h}) = \int_{- \infty}^\infty \left( \hat{F}_{d,h}(y | x_{d, h}) - \mathbb{1}_{\{y_{d,h} \leq y\}}\right)^2 \mathrm{d}y. $$
Interestingly, the CRPS can be reformulated (to a multiplicative constant) as :
$$CRPS(\hat{F}_{d,h}, y_{d, h}) = \int_{0}^1 \rho_\alpha \left(y_{d,h} , \hat{F}^{-1}_{d,h}(\alpha)\right) \mathrm{d}\alpha, $$
where $\hat{F}^{-1}_{d,h}(\alpha)$ actually corresponds to the predicted value at quantile $\alpha$. By approximating this integral as a Riemann sum, we can transform pinball scores over multiple quantiles into one single metric.

\begin{figure*}[!htb]
 \centering
 \includegraphics[width=0.8\textwidth]{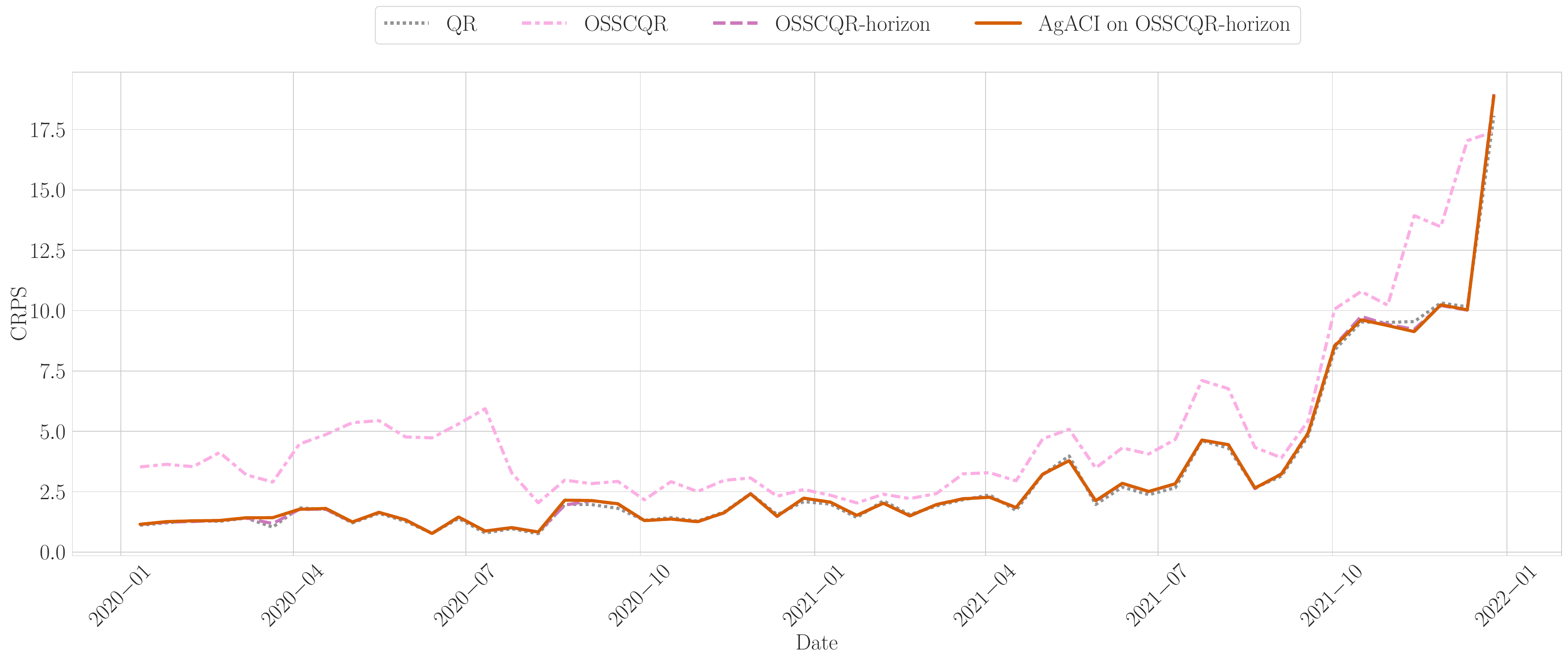}
 \caption{PIs's CRPS with different levels of conformalisation on the \textbf{quantile linear model}, depending on the time. The colors and shapes are associated with the conformalisation layers.}
 \label{fig:lin_crps}
\end{figure*}

\begin{figure*}[!htb]
 \centering
 \includegraphics[width=0.8\textwidth]{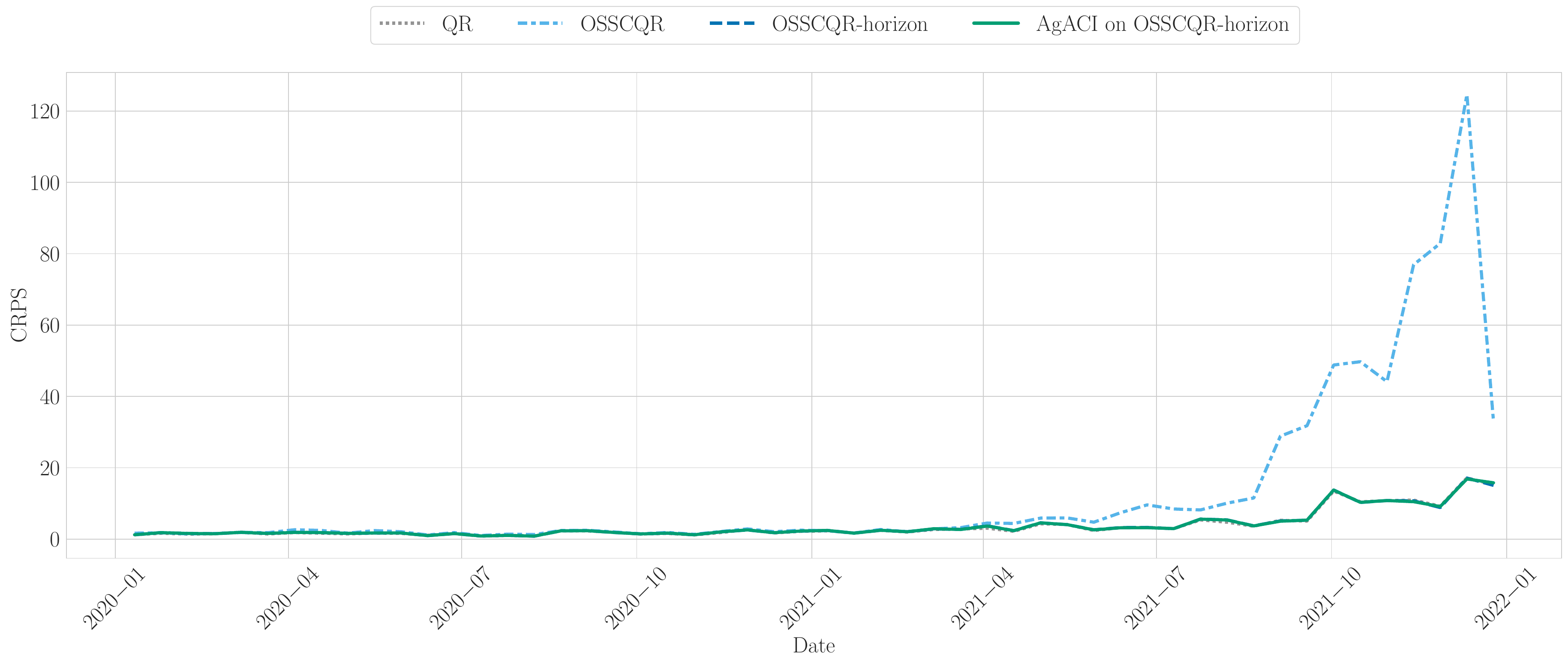}
 \caption{Same caption than \Cref{fig:lin_crps} but for the \textbf{quantile random forest model}.}
 \label{fig:qrf_crps}
\end{figure*}

\begin{figure*}[!htb]
 \centering
 \includegraphics[width=0.8\textwidth]{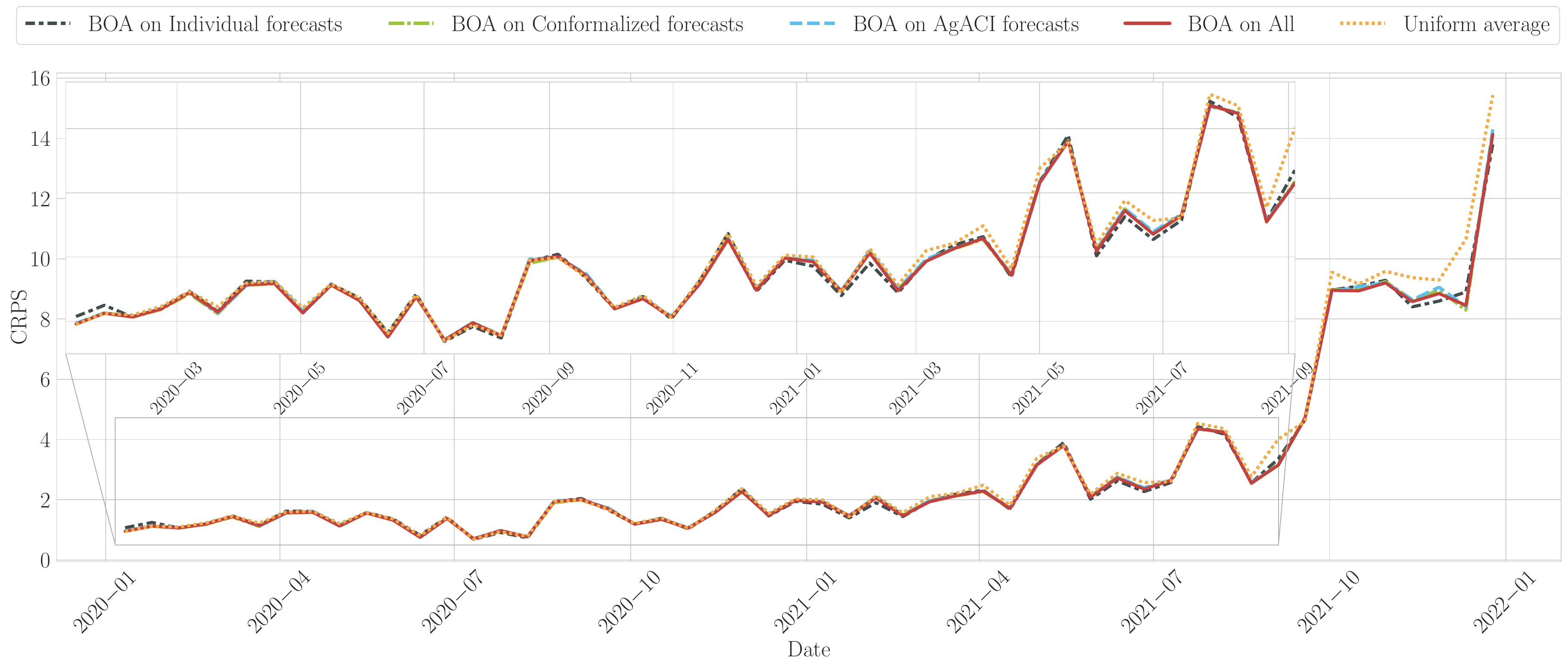}
 \caption{PIs's CRPS of online aggregation on multiple set of experts with windowing, depending on the time. The colors and shapes are associated with the set of experts.}
 \label{fig:agg_crps}
\end{figure*}

\end{document}

%% file: preamble.tex
\usepackage[round]{natbib}
\usepackage{dsfont}
\usepackage{enumitem}
\usepackage{microtype}
\usepackage{amsmath,amsthm,amssymb}
\usepackage[colorlinks]{hyperref}
\usepackage{mathrsfs}
\usepackage[noabbrev]{cleveref}
\usepackage{hhline}
\usepackage{booktabs}
\usepackage{soul}
\usepackage[dvipsnames]{xcolor}
\usepackage{graphicx}

\usepackage[utf8]{inputenc}
\usepackage{fourier} 
\usepackage{array}
\usepackage{makecell}

\usepackage{dblfloatfix}

\usepackage{cancel}
\usepackage{comment}

\definecolor{blindorange}{RGB}{215,131,37}
\definecolor{blindblue} {RGB}{81,172,226}
\definecolor{blindpurple} {RGB}{193,114,177}
\definecolor{blindgreen} {RGB}{33,145,106}
\definecolor{blindpink} {RGB}{252,174,226}
\definecolor{blindbrown} {RGB}{203,145,100}

\def\Tr{\rm{Tr}}
\def\Cal{\rm{Cal}}